\documentclass[12pt]{article}
\usepackage{graphicx}
\usepackage{amsmath}

\newcommand\pubnumber{}
\newcommand\pubdate{\today}

\def\napoli{Department of Statistics\\
Northwestern University, Evanston, USA}

\def\Title#1{\begin{center} {\Large #1 } \end{center}}
\def\Author#1{\begin{center}{ \normalsize #1 } \end{center}}
\def\Address#1{\begin{center}{ \it #1} \end{center}}

\newcommand\pubblock{\rightline{\begin{tabular}{l} \pubnumber\\
         \pubdate  \end{tabular}}}
\newenvironment{Abstract}{\begin{quotation}  }{\end{quotation}}

\def\Acknowledgements{\bigskip  \bigskip \begin{center} \begin{large}
             \bf ACKNOWLEDGEMENTS \end{large}\end{center}}

\def\beq{\begin{equation}}
\def\eeq#1{\label{#1}\end{equation}}
\def\eeqn{\end{equation}}

\def\beqa{\begin{eqnarray}}
\def\eeqa#1{\label{#1}\end{eqnarray}}
\def\eeqan{\end{eqnarray}}

\let\bar=\overbar

\def\Dslash{\not{\hbox{\kern-4pt $D$}}}
\def\dslash{\not{\hbox{\kern-2pt $\del$}}}

\def\msb{{\bar{\ssstyle M \kern -1pt S}}}

\begin{document}
\begin{titlepage}
\pubblock

\vfill
\Title{Community Detection with Node Attributes and Its Generalization}
\vfill
\Author{Y. Li}
\Address{\napoli}
\vfill
\begin{Abstract}
Community detection algorithms are fundamental tools to understand organizational principles in social networks. With the increasing power of social media platforms, when detecting communities there are two possible sources of information one can use: the structure of social network and node attributes. However structure of social networks and node attributes are often interpreted separately in the research of community detection. When these two sources are interpreted simultaneously, one common assumption shared by previous studies is that nodes attributes are correlated with communities. In this paper, we present a model that is capable of combining topology information and nodes attributes information without assuming correlation. This new model can recover communities with higher accuracy even when node attributes and communities are uncorrelated. We derive the detectability threshold for this model and use Belief Propagation (BP) to make inference. This algorithm is optimal in the sense that it can recover community all the way down to the threshold. This new model is also with the potential to handle edge content and dynamic settings.
\end{Abstract}
\vfill

\vfill
\end{titlepage}
\def\thefootnote{\fnsymbol{footnote}}
\setcounter{footnote}{0}

\section{Introduction}
Community detection is one of the critical issues when understanding social networks. 
In many real-world networks (e.g. Facebook, Twitter), in addition to topology of social network, content information is available as well. Even though different sources of information about social networks can be collected via social media, node attributes and the structure of networks are often interpreted separately in the research of community detection. Usually the primary attention of algorithms has only focused on the topology of the social networks while on the other hand, the decision of community assignments has been made solely based on node attributes. The partial use of data is tremendously inefficient. Sometimes, especially when the network is sparse, algorithms which are incapable of incorporating multiple data sources are often paralyzed and unsuccessful in recovering community assignment. It is of great interests to study how to incorporate topology features and node attributes into one algorithm.

Several papers address community detection with node attributes under the assumption that the observed node attributes are highly correlated with communities. The two main approaches are: heuristic measure-based models and probabilistic inference-based models. The heuristic measure-based models combine topology structure and node attributes in a heuristic function. L. Akoglu et al.~\cite{L.Akoglu} proposed a parameter-free identification of cohesive subgroups (PICS) in attributed graphs by minimizing the total encoding costs.Y. Zhou et al.~\cite{Y.Zhou} proposed SA-Cluster based on structural and attribute similarities through a unified distance measure.The probabilistic inference-based approach usually assumes that the networks are generated by random processes and uses probabilistic generative models to combine both topology and attributes. 
J. Yang et al.~\cite{J.Yang} developed Communities from Edge Structure and Node Attributes (CESNA) for detecting overlapping networks communities with node attributes.
In CESNA model, the links are generated by process of BigCLAM and node attributes can be estimated by separate logistic models.
B.F. Cai  et al.~\cite{B.F.Cai} proposed a popularity-productivity stochastic block model with a discriminative framework (PPSB-DC) as the probabilistic generative model.
Y.Chen et al.~\cite{Y.Chen} adopted Bayesian method and developed Bayesian nonparametric attribute (BNPA) model.
A nonparametric method was introduced to determine the number of communities automatically. 
These probabilistic generative models can be further categorized based on two different ways of modeling the stochastic relationship between attributes X, communities F and graph G. CESNA and BNPA assume that communities “generate” both the network as well as attributes (Figure 1 (c) ) however PPSB-DC assumes that communities can be predicted based on attributes and then network are generated based on communities (Figure 1 (d) ).

Even though many studies have shown that social ties are not made random but constrained by social position ~\cite{Mc}~\cite{Centola}, it is possible that the observed node attributes may not contribute much to social position so that they are uncorrelated with communities. When communities and node attributes are not correlated, adding nodes attributes into the above models will not give more information about communities. In this paper we propose an approach that allows us to go beyond the similarity between communities and node attributes.
One assumption we rely on is that node attributes will lead to heterogeneity in the degree of nodes (Figure 1 (e)). The idea of including heterogeneity in the degree in SBM was first introduced by Wang and Wong~\cite{Wang} and later revisited by Karrer~\cite{Karrer}.
By including this heterogeneity, our approach is able to solve the more challenging problem where node attributes and communities are uncorrelated. Our intuition is that the node attributes label not only the nodes but the edges as well. Due to heterogeneity in the degree, different types of edges carry different information of communities, therefore our approach should be able to recover the communities more accurately.

Another important problem of interests is to understand to which extend the extra information of node attributes will improve the performance, especially when communities and node attributes are not correlated. Here we are focusing on the detectability threshold for our new model. E. Mossel et al.~\cite{Mossel1} have proven that there exists a phase transition in the detectability of communities for two equal size communities  in stochastic block model. S. Heimlicher et al.~\cite{Heimlicher} investigated the phase transition phenomena in more general context of labelled stochastic block model and generalized the detectability threshold. A. Ghasemian et al.~\cite{Ghasemian} derived the detectability threshold in dynamic stochastic block model as a function of the rate of change and the strength of the communities. In this paper we derive the detectability thresholds for community structure in stochastic block model with node attributes and compare it with the original thresholds while no information of node attributes is available.

 \begin{figure}[htb]
 	\centering
 	\includegraphics[scale=0.4]{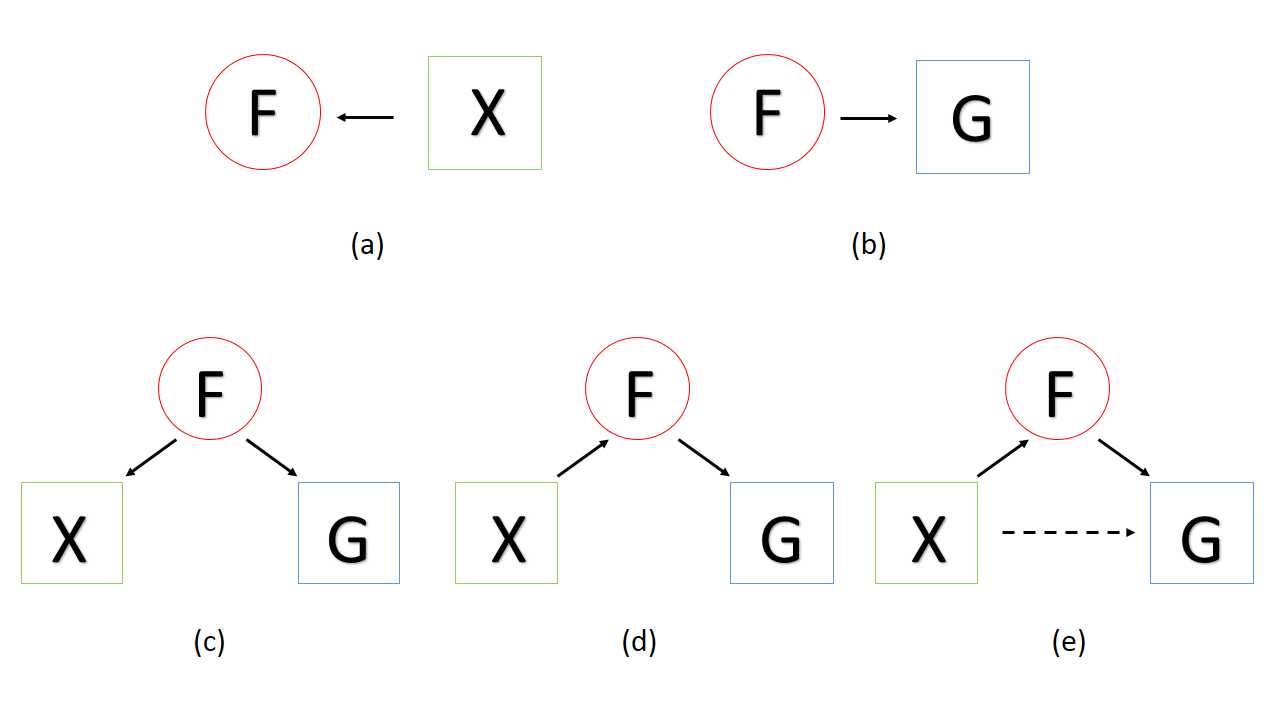}
 	\caption{Ways of modeling the stochastic relationship between attributes X, communities F and graph G. Circles represent latent community assignment and squares represents observed variables.}
 \end{figure}

\section{Model}
The stochastic block model (SBM) is a classic probabilistic generative model of community structure in static networks~\cite{Holland}~\cite{Faust}~\cite{Snijders}. Here, we develop a generative model by extending SBM to include heterogeneity due to node attributes in the degree of nodes. 
 In our model, we first assign nodes with different nodes attributes to different communities and then generate the topology of network based on both the community assignment and the node attributes (Figure 1 (e)). The graphical model in Figure 1 (e) can be seen as an extension of the graphic model in Figure 1 (d). The main reason for generalizing the graphic model in Figure 1 (d) instead of the graphic model in Figure 1 (c) is that the graphic model in Figure 1 (d) is a combination of graphical models in Figure 1 (a) and Figure 1 (b), which are corresponding graphical models for clustering problem and community detection in stochastic block model. Therefore we find the graphic model in Figure 1 (d) is a better candidate to combine topology information and node attributes information.
In our model, we also assume that all the node attributes are categorical variables. 
Finally we correct the degree of nodes based on node attributes, which leads to sub-communities structure (Figure 2). This assumption allows heterogeneity in communities and generalize the community in SBM.

\begin{figure}[htb]
	\centering
	\includegraphics[scale=0.6]{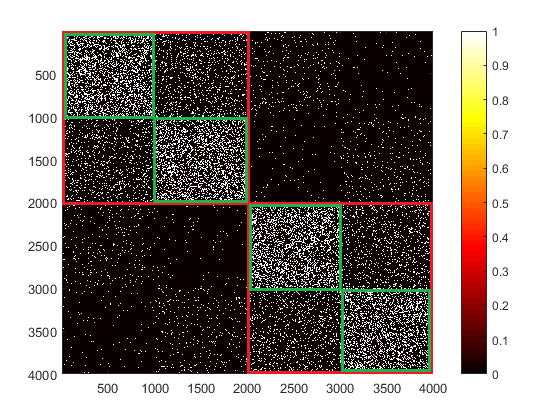}
	\caption{heat map of block matrix, red squares represent two primary communities, green squares represent sub-communities in primary communities.}
\end{figure}

We formally describe the generative process of a graph $G=\{V,E,x_{1},x_{2},\dots,x_{m}\}$ under stochastic block model with node attributes, where $x$ represents attributes, as follows. 
First, we construct an one-to-one map of node attributes from m-dimensional point $ \{x_{1},x_{2},\dots,x_{m}\}$ to 1-dimensional point $\{X_{r}\}$, where $m$ is the number of different types of observed attributes and $r$ is from $1$ to $R$. Then we assign each of the n nodes $i \in V$ into $R$ group according to node attributes and the number of nodes in each group is $n_{r}$. Using a prior $q_{k,r}$, we assign $n_{r}$ nodes in attributes category $r$ into K communities. We then generate the $(i,j)$th element in adjacency matrix $A$ ccording to a Bernoulli distribution with probability $P_{\{k_{i},r_{i}\},\{k_{j},r_{j}\}}$, where $ k_{i}$ is the community assignment for node $i$, $r_{i}$ is the attributes category for node $i$ and $P_{\{k_{i},r_{i}\},\{k_{j},r_{j}\}}$ is the probability of forming an edge between a node from community $ k_{i}$ with attributes $X_{r_{i}}$ and a node from community $ k_{j}$ with attributes $ X_{r_{j}}$.The full likelihood of graph under SBM with node attribute is:  
\begin{equation}
P(E,k|X,P)= (\prod_{i} q_{k_{i},r_{i}})(\prod_{i,j \in E} P_{\{k_{i},r_{i}\},\{k_{j},r_{j}\}} \prod_{i,j \notin E}(1- P_{\{k_{i},r_{i}\},\{k_{j},r_{j}\}}))
\end{equation}
Since $ P_{\{k_{i},r_{i}\},\{k_{j},r_{j}\}}=O(\frac{1}{n})$, sometimes it’s easier to work with the rescale matrix $c_{\{k_{i},r_{i}\},\{k_{j},r_{j}\}}$. When two nodes are from group $K_{1},K_{2}$ with category of attributes $a,b$, the rescale matrix $c_{\{K_{1},a\},\{K_{2},b\}}=nP_{\{K_{1},a\},\{K_{2},b\}}$.
For subsequent analysis, we will focus on the choice of uniform prior $q_{k,r}= \frac{1}{K} $  since we are interested in the detectability threshold when attributes are not correlated with communities. We will also limit ourselves to an algorithmically difficult case of block model, where every group k has the same average degree conditional on the type of edge:
\begin{equation}
c_{{a}{b}}=\frac{n_{b}}{K^2} \sum_{k_{1}}\sum_{k_{2}} P_{\{k_{1},a\},\{k_{2},b\}}= \frac{n_{b}}{K} \sum_{k_{2}} P_{\{k_{1},a\},\{k_{2},b\}}   \text{    for any } k_{1}.
\end{equation}
If this is not the case, reconstruction can be achieved by labeling nodes based on their degrees.

\section{Detectability threshold in SBM with node attributes}
The best-known rigorous detectability threshold in sparse SBM has been derived by E. Mossel et al.~\cite{Mossel1}. In the sparse partition model, where $p=\frac{a}{n}$,$q=\frac{b}{n}$ and $a>b>0$ , the clustering problem is solvable in polynomial time if $(a-b)^2>2(a+b)$. However for $K\ge3$ it is still an open question to find a rigorous detectability threshold in SBM. The Kesten-Stigum (KS) threshold in statistical physics can be treated as a non-rigorous threshold for $K\ge3$~\cite{KS1}~\cite{KS2}. Let $G$ be generated by SBM$(n,k,a,b)$ and define $SNR=\frac{|a-b|}{\sqrt{k(a+(k-1)b)}}$. If $SNR>1$ then the clustering problem is solvable and the Kesten-Stigum (KS) threshold can be achieved in polynomial time. In the sparse regime, $|E|=O(n)$, the graph generated by SBM is locally treelike in the sense that all most all nodes in the giant component have a neighborhood which is a tree up to distance $O(log(n))$. Therefore the threshold for reconstruction on tree can provide good insight into reconstruction on SBM.

As mentioned before, one intuition is that node attributes label the edges, therefore we consider a multi-type branching process of edges to generate the tree that approximates the graph generated by SBM with node attributes. By defining a Markov chain on the infinite tree $T=(V,E,X)$, we can derive the construction thresholds on SBM with node attributes.

To construct the multi-type branching process, we first label the edge by the categories of node attributes at the two ends of the edge as $L\{X_{a},X_{b}\}$, where $X_{a}$ is the attributes for the node that is closer to the root, $X_{b}$ is the attributes for the node at the far-end and $a,b$ is from $1$ to $R$. So there are $R^2$ different types of edges. Map $\{X_{a},X_{b}\}$ to $(a-1)*R+ b$ and relabel the $L\{X_{a},X_{b}\}$ type edge as $L\{(a-1)*R+ b\}$.
Let $R^2*R^2$ dimensional matrix C be the matrix describing the expected number of children, where $c_{ij}$ is the expected number of type $ L\{i\}$ edges induced by one type $L\{j\}$ edge. Note that one $L\{X_{a_{1}},X_{b_{1}}\}$ type of edge will give birth to $L\{X_{a_{2}},X_{b_{2}}\}$type of edges if and only if $ b_{1}=a_{2}$. Let $x= [\frac{i-1}{R}]+1$ and $y= i-[\frac{i-1}{R}]$ and $ z= j-[\frac{j-1}{R}]$,

\begin{equation}
c_{ij} = 
\begin{cases}
0 & \text{if } x\not=z,\\
c_{xy}&\text{if }  \text{otherwise}.
\end{cases}
\end{equation}

When moving outward a type $L\{X_{a},X_{b}\}$ edge, the $K*K$ stochastic transition matrix $\sigma$ associate with the edge can be defined as:
\begin{equation}
\sigma^{k_{1}k_{2}}_{ab}=\frac{\frac{n_{b}}{K} P_{\{k_{1},a\},\{k_{2},b\}}}{c_{ab}}.
\end{equation}
The largest eigenvalue for the $K*K$ stochastic transition matrix $\sigma$ is 1 and let the second largest eigenvalue be $\lambda_{ab}$. 
Define $m_{ij}$ in the $R^2*R^2$ matrix $M_{1}$ as $ c_{ij}*\lambda_{[\frac{i-1}{R}]+1, i-[\frac{i-1}{R}]}^2$. The robust reconstruction is possible when the absolute value of largest eigenvalue for matrix $M_{1}$ exceeds $1$~\cite{Ghasemian}\cite{Mossel2}.
 
\section{Belief propagation }
To recover the community assignments in SBM with node attributes, we use Bayesian inference to learn the latent community:

\begin{equation}
P(k|E,X,P)= \frac{ P(k,E|X,P)} {\sum_{g}P(E|g,X,P)},
\end{equation}
where $k$ is the estimated group assignment and $g$ is the original group assignment. The distribution is too complex to compute directly since$\sum_{g} P(E|g,X,P)$ runs over exponential number of terms. In the regime $|E|=O(n)$, the graph is locally treelike therefore belief propagation, which is exact to calculate the marginal probability of community assignment on a tree, can be applied to calculate Bayesian inference efficiently. We’ll show that BP is an optimal algorithm in the sense that it can reach the detectability thresholds for SBM with node attributes. 

To write the belief propagation equation, we define conditional marginal probability, denoted as $\psi_{k_{i}}^{i\to j}$, which is the probability that node $i$ belongs to group $k_{i}$ in the absence of node {j}. We can compute the messenger from $i$ to $j$ as:

\begin{equation}
\psi_{k_{i}}^{i\to j}=
\frac{1}{Z^{i\to j}} q_{k_{i}r_{i}}\prod_{l\in \partial i \setminus j} 
[\sum_{k_{l}} c_{\{k_{l},r_{l}\},\{k_{i},r_{i}\}}^{A_{il}}
(1- \frac{c_{\{k_{l},r_{l}\},\{k_{i},r_{i}\}}^{1-A_{il}}}{n})\psi_{k_{i}}^{l\to i}],
\end{equation}
where $ A_{il}$ is the $(i,l)$th element in the adjacency matrix for the graph generated by SBM with node attributes, $\partial i$ denotes all the nodes connected to $i$ and $ Z^{i\to j}$ is a normalization constant ensuring $\psi_{k_{i}}^{i\to j}$ to be a probability distribution. The marginal probability $\psi_{k_{i}}^{i}$ can be calculated as:
\begin{equation}
\psi_{k_{i}}^{i}=
\frac{1}{Z^{i}} q_{k_{i}r_{i}}\prod_{l\in \partial i } 
[\sum_{k_{l}} c_{\{k_{l},r_{l}\},\{k_{i},r_{i}\}}^{A_{il}}
(1- \frac{c_{\{k_{l},r_{l}\},\{k_{i},r_{i}\}}^{1-A_{il}}}{n})\psi_{k_{i}}^{l\to i}],
\end{equation}
where $ Z^{i}$ is a normalization constant ensuring $\psi_{k_{i}}^{i}$ to be a probability distribution. 
In SBM with node attributes, we have interactions between all pairs of nodes, therefore we have $n(n-1)$ messengers to update for one iteration. To reduce the computational complexity to $O(n)$, we follow past work on BP for SBM\ cite{Decelle}. At the cost of making $O(\frac{1}{n})$ approximation to the messenger, when there is no edge between $i$ and$j$, the messenger from $i$ to $j$ can be calculated as:
\begin{equation}
\psi_{k_{i\to j}}^{i}=\psi_{k_{i}}^{i}.
\end{equation}
Now only messengers sent on edges are needed to be calculated. By introducing an external field, the messenger from $i$ to $j$ when there is an edge between $i$ and $j$ can be approximated as: 
\begin{equation}
\psi_{k_{i}}^{i\to j}=
\frac{1}{Z^{i}} q_{k_{i}r_{i}}e^{-h_{k_{i}r_{i}}}\prod_{l\in \partial i }
[\sum_{k_{l}} c_{\{k_{i},r_{i}\},\{k_{l},r_{l}\}}
\psi_{k_{i}}^{l\to i}],\label{fixpoint}
\end{equation}
where the external field $ h_{k_{i}r_{i}}$ can be defined as:
\begin{equation}
h_{k_{i}r_{i}}=\frac{1}{n}\sum_{l}\sum_{k_{l}}
c_{\{k_{i},r_{i}\},\{k_{l},r_{l}\}}\psi_{k_{l}}^l.
\end{equation}
It’s worth noting that $ \psi_{k_{i}}^{i\to j}= q_{k_{i}r_{i}}$ is a fixed point in \eqref{fixpoint}.

\section{Phase transition in BP and simulation}
In this section, we will study the stability of the fixed point under random perturbations. As discussed above, in the sparse regime, the graph generated by SBM with node attributes is locally treelike. Here consider such a tree with $d$ levels. On the leave $m_{d}$ the fixed point is perturbed as $\psi_{k_{m_{d}}}^{m_{d}}= q_{k_{m_{d}}r_{m_{d}}} +\epsilon_{k_{m_{d}}}^{m_{d}}$, where $\epsilon_{k_{m_{d}}}^{m_{d}}$ is $i.i.d.$ random variable.
Then the influence of perturbation on leave $m_{d}$ to the root $m_{0}$ can be calculated as:
\begin{equation}
\epsilon^{m_{0}}=\prod_{\{{a}{b}\}} T_{ab}^{d_{ab}}\epsilon^{m_{d}},
\end{equation}  
where $d_{ab}$ is the number of type $L\{X_{a},X_{b}\}$ edges on the path from leave $m_{d}$ to the root $m_{0}$ and $ T^{ab}$ is the transfer matrix for type $L\{X_{a},X_{b}\}$ edges, which, by following the calculation in \cite{Decelle}, can be defined as:
\begin{equation}
T_{ ab}^{k_{1}k_{2}}=q_{{k_{1}}a}(k\sigma^{k_{1}k_{2}}_{ab}-1).
\end{equation}
As $d\to \infty$, $d_{ab} \to \infty$ as well,therefore $\epsilon^{m_{0}}\approx\prod_{all\{ab\}} \upsilon_{{a}{b}}^{d_{ab}}\epsilon^{m_{d}} $,where $\upsilon_{{a}{b}}$ is the largest eigenvalue for $ T^{ ab}$.
Now let us consider the variance at root $m_{0}$ induced by the random perturbation on all leaves at level $d$. Since the influence of each leaf is independent, the variance of the root can be written as: 
\begin{equation}
Var(\epsilon^{m_{0}})=\sum_{\text {all the path } (r\sim m_{d})}
\prod_{{\{{a}{b}\}}}\upsilon_{{a}{b}}^{2d_{ab}}	Var(\epsilon^{m_{d}}).\label{branching}
\end{equation}

When the variances on leaves are amplified exponentially, the fixed point is unstable and BP algorithm is able to recover the community assignment with high probability, otherwise the perturbation on leaves will vanish and the fixed point in stable under BP algorithm. From eq.\eqref{branching}, when $\epsilon^{m_{d}}$ is $i.i.d.$, to determine the phase transition in BP, it's sufficient to calculate $Z_{d}=\sum_{all the path (r\sim m_{d})}
\prod_{{all\{{a}{b}\}}}\upsilon_{{a}{b}}^{2d_{ab}}
$. This calculation can be done by viewing this summation as a weight associated multi-type branching process.   
Consider thus a multi-type branching process with Possion distribution with mean $c_{ab}$ if the parent-child edge in the corresponding multi-type branching process belongs to type $L\{X_{a},X_{b}\}$. The variance amplified along the tree generated by the above multi-type branching process and the expected values of the variance at level $d$ can be calculated as:
\begin{equation}
E(Z_{d}|m_{0})=\boldsymbol{1}^{T}M_{2}^de_{m_{0}},
\end{equation}
where the $(i,j)$th element of $M_{2}$ is $c_{ij}\upsilon_{{[\frac{i-1}{R}]+1},{i-[\frac{i-1}{R}]}}^2$, $e_{m_{0}}$ is an $R^2$-dimensional unit vector with the $r$th element equal to $1$ and $r$ is the node attribute type of the root node $m_{0}$.
When the largest eigenvalue of $M_{2}$ exceeds $1$, the fixed point of BP is unstable and the community is detectable. Noting that $\lambda_{ab}=\upsilon_{ab}$, therefore BP is an optimal algorithm in the sense that it can reach the detectability threshold in SBM with node attributes even when node attributes and communities are not correlated.

Next, we compare the detectability thresholds for SBM with node attributes with the detectability threshold for the original SBM without information of node attributes. In the following discussion, we will limit ourselves to the case where $n_{r}=\frac{n}{R}$, $q_{k_{i},r_{i}}=\frac{1}{K}$ and $ C_{\{k_{i},r_{i}\},\{k_{j},r_{j}\}}$ satisfy equation \eqref{simulation},
\begin{equation}
	c_{\{k_{i},r_{i}\},\{k_{j},r_{j}\}}=
	\begin{cases}
		a & \text{if } k_{i}= k_{j} \text{and } r_{i}= r_{j} \\
		b &\text{if }  \ k_{i}= k_{j} \text{and } r_{i}\not= r_{j} \\
		c &\text{if } \text{otherwise},
	\end{cases}
	\label{simulation}
\end{equation}
where $ a\geq b\geq c$. For SBM with node attributes, the community is detectable if
\begin{equation}
	\xi_{1}=\frac{(a-c)^2}{a+(K-1)c}+(R-1)\frac{(b-c)^2}{b+(K-1)c}>KR.
\end{equation}
For SBM without information of node attributes, the community is detactable if
\begin{equation}
	\xi_{2}=\frac{(a+(R-1)b-Rc)^2}{a+(R-1)b+(K-1)Rc}>KR.
\end{equation}
By simple calculation, it can be shown that $\xi_{1}\ge\xi_{2}$, therefore even in the situation where the observed node attributes are uncorrelated with communities, including node attributes into model will give us more infomation about communities.

We conduct the following simulation to verify the claim of phase transition in BP.
Considering for simplicity only two communities and two node attributes, we generate a series of graphs by SBM with node attributes for $4000$ nodes and various choice of $(a,b)$ when controlling average degree to be $5$. We use $\eta=
\frac{a}{b}$ to represnt different choices of $(a,b)$ and $\epsilon=\frac{c}{b}$ to represent the strength of communities. When  $\epsilon=0$ the clusterings are maximally strong while at $\epsilon=1$ the clusterings are weak.
The accuracy of reconstruction is measure by $overlap$ matric introduced by \cite{Decelle}.
 \begin{figure}[htb]
 	\centering
 	\includegraphics[scale=0.8]{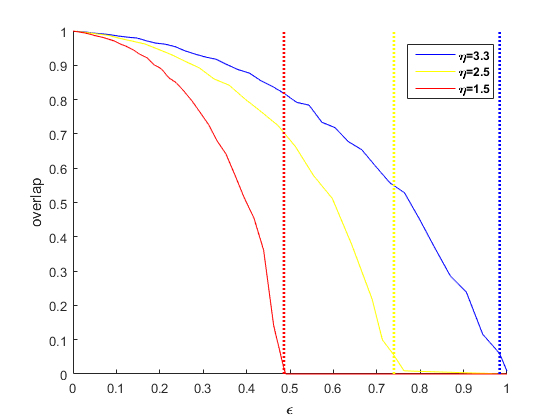}
 	\caption{Overlap as a function of $\epsilon$ for various values of $\eta$. Dash lines mark the theoretical detectability thresholds for the choice of $(\epsilon,\eta)$.}
 \end{figure}

 In figure 3, we plot $overlap$ metric against $\epsilon$ for different values of $\eta$ and for each curve, we use a vertical dash line in the same color as the corresponding curve to indicate the detectability threshold. Figure 3 shows that BP can recover communities that are positively correlated with true communities all the way down to the detectability thresholds for various choice of $(\epsilon, \eta)$. The algorithm has larger $overlap$ metric with smaller $\epsilon$.
\section{Conclusion}
In this paper, we consider a model that uses information of nodes attributes in a different way such that this approach will provide more information of latent communities beyond the information carried by SBM even when node attributes are not correlated with communities. We have derived a theoretical detectability threshold for SBM with node attributes, which coincides with phase transition in BP. We also conduct a numerical analysis of the phase transition in BP. While constricted to the two symmetric communities with two node attributes, this condition is sufficient to illustrate how the information of node attributes affects detectability even the node attributes are not correlated with communities. 

A nature extension will include edge contents and dynamic settings into the model. Our approach can be applied to this case by including different type of edges into the multi-branching process. On the theoretical front, it has been conjectured~\cite{ Mossel1} that, for $K \ge 5$, there’s a regime that the clustering problem is solvable but not in polynomial time. Emmanuel Abbe and Colin Sandon~\cite{Abbe} have developed a non-efficient algorithm that is shown to break down KS threshold at $K=5$ in SBM. As a future work, we’ll try to develop an algorithm that can break down the detectability threshold in our model for large numbers of groups.

\Acknowledgements
I am grateful to Professor Wenxin Jiang and Professor Noshir Contractor for helpful discussion.

\end{document}